\documentclass[Journal]{IEEEtran}
\usepackage{graphicx, epstopdf}
\usepackage{amssymb, amsmath}
\usepackage{commath}
\usepackage{setspace}
\usepackage{acronym}
\usepackage{mathrsfs}
\usepackage{ifthen}
\usepackage{textcomp}
\usepackage{float}
\usepackage{subfigure}
\usepackage{color}
\usepackage{cite}

\IEEEoverridecommandlockouts

\begin{document}
\bstctlcite{ICC09_Ref2:BSTcontrol}

\title{Molecular MIMO Communication Link}

\author{\authorblockN{Changmin Lee, Bonhong Koo, Na-Rae Kim, Birkan Yilmaz,\\ Nariman Farsad\textsuperscript{\dag}, Andrew Eckford\textsuperscript{\dag}, and Chan-Byoung Chae} \\
\authorblockA{{
Yonsei Institute of Convergence Technology, School of Integrated Technology, Yonsei University, Korea \\
\textsuperscript{\dag}Dept. of Electrical Engineering and Computer Science, York University, Toronto, Canada \\
Email: cbchae@yonsei.ac.kr} } }

\maketitle

\begin{abstract}
In this demonstration, we will present the world's first molecular multiple-input multiple-output (MIMO) communication link to deliver two data streams in a spatial domain. We show that chemical signals such as concentration gradients  could be used in MIMO fashion to transfer sequential data. Until now it was unclear whether MIMO techniques, which are used extensively in modern radio communication, could be applied to molecular communication. In the demonstration, using our devised MIMO apparatus and carefully designed detection algorithm, we will show that we can achieve about 1.7 times higher data rate than single input single output (SISO) molecular communication systems. 

\end{abstract}

\section{Background}

Molecular communication is a biologically inspired form of communication, where chemical signals are used to transfer information \cite{Farsard_arXiv14}. Molecular communication could be used in places where radio based communication fails or is inefficient: for example,  city infrastructure  monitoring in smart cities at macroscale~\cite{Wassell10}, and body area nanonetworks for health monitoring and targeted drug delivery at microscale~\cite{ata12CM}.

Most previous work on molecular communication has focused on microscale systems and nanonetworks such as diffusion-based intra- and inter-cell communications \cite{lla12WC}. Most of these works have been theoretical, and only recently the have been experimental implementations of molecular communication systems \cite{Farsad13}, where reliable communication was achieved. There have also been a number of attempts at mimicking pheromone-based communication \cite{col09}. Although these systems were not designed for transferring sequential data. More recently, it was demonstrated that the nonlinearity in~\cite{Chae_JSAC14} could be modelled as noise. 

In our prior work, the world's first macro scale molecular communication link \cite{Farsad13} was demonstrated at IEEE INFOCOM~2014 \cite{infocom_demo}. In this demonstration we will show the world's first molecular multiple-input multiple-output (MIMO) communication link, where the transmitter and the receiver are equipped with multiple sprays and sensors to further increase the data rate. MIMO is a technique, which is used in modern radio communication to increase transmission data rate. The feasibility of using MIMO in molecular communication, however, has not been demonstrated in the past. In our MIMO design, we implement our own signal detection algorithms that are different from classical RF MIMO communication. The algorithms will be described in more detail in an extended draft.

\section{Test-bed and Demonstration Output}


\subsection{Hardware Layout}

\begin{figure}[t]
	\begin{center}
		\subfigure[Transmitter]{\includegraphics[width=0.27\textwidth,keepaspectratio]{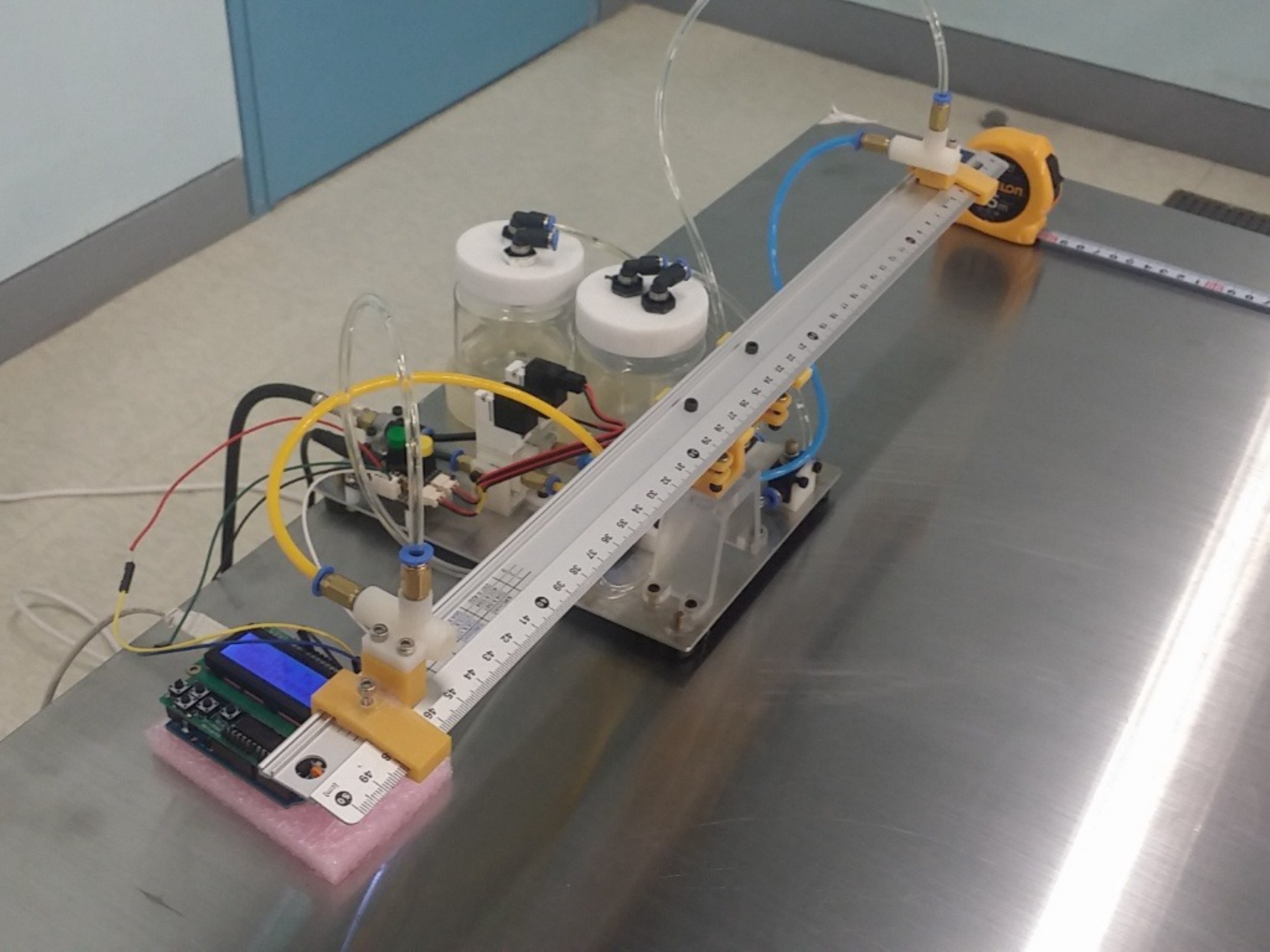}} %
		\subfigure[Receiver]{\includegraphics[width=0.18\textwidth,keepaspectratio]{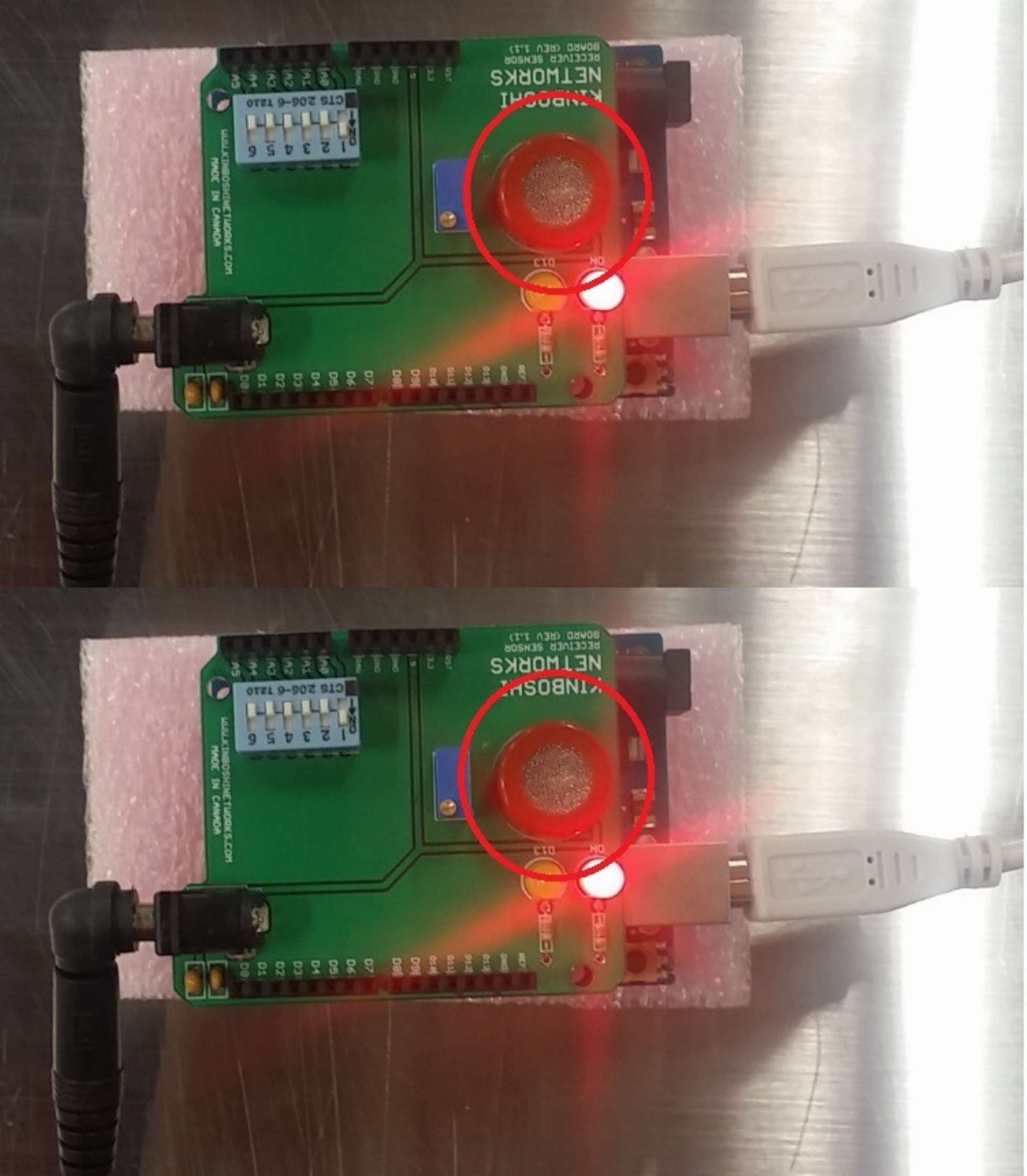}} %
	\end{center}
	\caption{The tabletop molecular MIMO communication platform.}
	\label{Fig:test_bed_mimo}
\end{figure}

Our system consists of a molecular MIMO transmitter and receiver as shown in Fig.~\ref{Fig:test_bed_mimo}. The propagation channel in between is several meters of free-space. The transmission consists of: 1) a simple user interface for text entry, 2) a microcontroller for executing transmitter algorithms, 3) two reservoir for chemicals, and iv) two chemical release mechanism (i.e. two sprays). At the receiver, the hardware consists of: 1) two chemical sensors for MIMO operation, 2) two microcontrollers that demodulate and decode the signal, and 3) a computer for displaying and visualizing results.

%
%

\label{MIMO test bed}

Although any sequential data could be transported by our setup, for this demonstration we will consider short string of text data. Text-based information is very important to sensor networks and command-based communication systems. The information delivery rate for the prior platform we demonstrated at IEEE INFOCOM 2014 was low because of inexpensive components and use of single type of chemical~\cite{infocom_demo}. This year, we utilize spatial domain to increase the transmission data rate. Unlike prior work in RF communication, non-coherent detection is required since the coherent time of the molecular channel is zero.



\subsection{Health and Safety}

We have the same demonstration conditions as described in~\cite{infocom_demo}. As part of the demonstration low volumes of alcohol is diffused in open air. There will not be any chemical risks since the alcohol used will be safe for human consumption and of a small quantity and concentration. Moreover, we will perform the demonstration behind a transparent shielded screen to respect religious sensitivities and avoid any unwanted alcohol odours in the conference venue. 


%



\section{Application}

\begin{figure}[t]
	\begin{center}
       \includegraphics[width=3.1in]{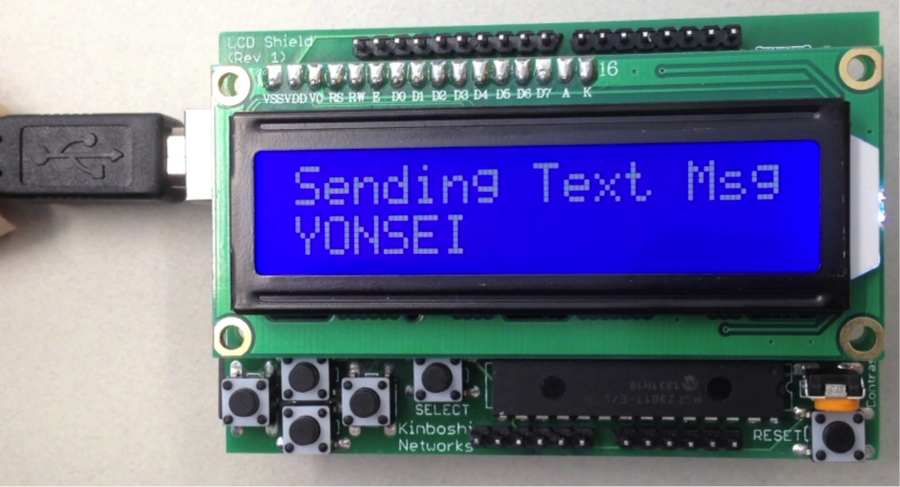}
	\end{center}
    \caption{Sample text message entered at the transmitter.}\label{Fig:sampleTxRx}
\end{figure}
\begin{figure}[!t]
	\centering
	\includegraphics[width=1.0\columnwidth,keepaspectratio]
	{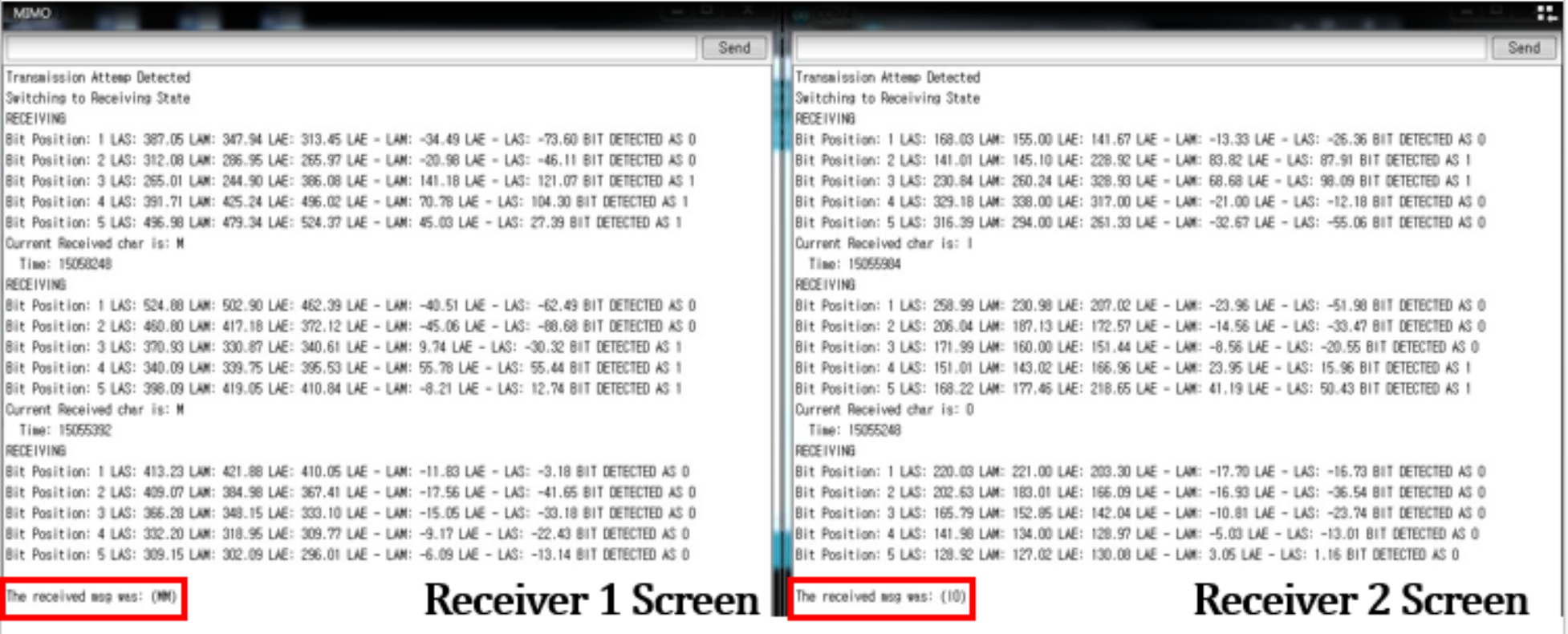}
	\caption{On the receiver screens the decoded characters are seen.}
	\label{Fig:receiver_screen}
\end{figure}

The main goal of the demonstration is to show that messages can be continuously and reliably carried via molecular MIMO. Fig.~\ref{Fig:sampleTxRx} shows the sample text entered at the transmitter and Fig.~\ref{Fig:receiver_screen} illustrates the receiver screen. As can be seen from the figure, we decode three alphabets at each receiver sensor. Table~I compares the transmission time and the data rate of SISO and MIMO systems, from which we observe that the MIMO system show 1.7 times higher data rate than the SISO system. The data rate enhancement is not exactly two times even though we use two sprays and two sensors. This is because of the need for interference compensation and the system overhead due to start and end of communication indicators. 

Most mathematical models developed for molecular communication have relied on Fick's diffusion equation and Monte-Carlo simulation~\cite{Yilmaz_simul}. Moreover, most prior work has assumed perfect transmission, propagation, and reception~\cite{Chae_JSAC13}. These assumptions, however, do not hold in practice, and more realistic models based on experimental data are necessary~\cite{Chae_JSAC14}. Thus, we believe that more accurate MIMO channel models can also be derived through using experimental data obtained from our MIMO platform. In terms of industrial interest, our platform could be extended towards structural high speed health monitoring (smart cities) applications, and for transmitting commands to robots in subterranean areas~\cite{Wassell10}. 

\section{Conclusion}
In this demonstration, we present the first macroscale molecular MIMO communication system that could reliably transmit short text messages. Our goal is to show that molecular communication can be used as an alternative to radio communication in challenging environments. To improve the low transmission rate of molecular communication, we implement novel molecular MIMO detection algorithms. The main challenge in our design was implementing a signal separation algorithm for the molecular MIMO channel, since MIMO detection algorithms for classical RF communications could not be directly applied. We hope to motivate researches, and fill a gap between theory and practice of molecular communication. \\



\section{Acknowledgement}
This research is funded by the MSIP (Ministry of Science, ICT \& Future Planning), under the ÔÔIT Consilience Creative ProgramÕÕ (NIPA-2014-H0201-14-1002) supervised by the NIPA (National IT Industry Promotion Agency) and by the Basic Science Research Program (2014R1A1A1002186) funded by the MSIP, through the National Research Foundation of Korea and by the ICT R\& D programme of MSIP/IITP. The authors would like to thank C. Kim for his help in implementing the hardware.

\begin{table}[t]
	\caption{Experiment results of the macro-scale SISO and MIMO molecular communication testbed.}
	\label{tab_comparison_testbed}
	\centering
	\begin{tabular}{c c c}
		\hline  
		Type                 & Transmission time~(s)      & Data rate (bps) \\
		\hline  
		SISO          & 108	     & 0.28       \\
		MIMO          & 63	     & 0.48       \\
		\hline
	\end{tabular}
\end{table}

\bibliographystyle{IEEEtran}
\bibliography{IEEEabrv,Ref}

\end{document}